\documentclass[12pt,preprint]{aastex}

\usepackage{graphicx}
\usepackage{subfig}
\usepackage{epstopdf}
\usepackage{verbatim}
\usepackage{multirow}
\usepackage{float}
\usepackage{graphics,epsfig}
\usepackage{natbib}
\usepackage{lscape}
\def\spose#1{\hbox to 0pt{#1\hss}}
\def\lta{\mathrel{\spose{\lower 3pt\hbox{$\mathchar"218$}}
     \raise 2.0pt\hbox{$\mathchar"13C$}}}
\def\gta{\mathrel{\spose{\lower 3pt\hbox{$\mathchar"218$}}
     \raise 2.0pt\hbox{$\mathchar"13E$}}}

\def\figure#1#2 {\par{\narrower\noindent {\bf Fig. #1}
   \hskip 2mm #2\par}\bigskip\noindent}
\def\table#1#2 {\par{\narrower\noindent {\bf Tab. #1}
   \hskip 2mm #2\par}\bigskip\noindent}

\def\registered{{\ooalign{\hfil\raise .00ex\hbox{\scriptsize R}\hfil\crcr\mathhexbox20D}}}

\usepackage{amsmath}
\usepackage{accents}
\newlength{\dhatheight}

\slugcomment{Version: \enskip \today}

\shorttitle{Habitable Zone of Kepler-16}

\shortauthors{Moorman et al.}

\begin{document}


\title{
The Habitable Zone of Kepler-16: \\
Impact of Binarity and Climate Models
}

\bigskip
\bigskip

\author{S. Y. Moorman$^1$, B. L. Quarles$^2$, Zh. Wang$^1$, and M. Cuntz$^1$}

\affil{$^1$Department of Physics, Box 19059}
\affil{University of Texas at Arlington, Arlington, TX 76019;}
\email{sarah.moorman@mavs.uta.edu; zhaopeng.wang@mavs.uta.edu; cuntz@uta.edu}

\affil{$^2$Homer L. Dodge Department of Physics and Astronomy}
\affil{University of Oklahoma, Norman, OK 73019;}
\email{billylquarles@gmail.com}

\bigskip

%%%%%%%%%%%%%%%%%%%%%%%%%%%%%%%%%%%%%%%%%%%%%%%%%%%%%%%%%%%%%%%%%%%%%%%%

\begin{abstract}
We continue to investigate the binary system Kepler-16, consisting of
a K-type main-sequence star, a red dwarf, and a circumbinary Saturnian
planet.  As part of our study, we describe the system's habitable zone based
on different climate models.  We also report on stability investigations
for possible Earth-mass Trojans while expanding a previous study
by B.~L. Quarles and collaborators given in 2012.  For the climate models
we carefully consider the relevance of the system's parameters.  Furthermore,
we pursue new stability simulations for the Earth-mass objects starting along
the orbit of Kepler-16b.  The eccentricity distribution as obtained prefers
values close to circular, whereas the inclination distribution remains flat.
The stable solutions are distributed near the co-orbital Lagrangian points,
thus enhancing the plausibility that Earth-mass Trojans might be able to exist
in the Kepler-16(AB) system.
\end{abstract}


\clearpage

%%%%%%%%%%%%%%%%%%%%%%%%%%%%%%%%%%%%%%%%%%%%%%%%%%%%%%%%%%%%%%%%%%%%%%%%

\section{Introduction}

Kepler-16 is a well-documented example of a closely separated binary
system with a Saturnian planet in a P-type orbit \citep{sla11,doy11}.
P-type orbit means that the planet encircles both stars instead of
only one star with the other star acting as a perturber \citep{dvo82}.
Previous results on the existence and orbital properties of planets in
binary systems have been given by, e.g., \cite{rag06,rag10} and
\cite{roe12}, among others.  Detailed information on the abundance
of circumstellar planets has been given by \cite{wan14} and \cite{arm14}.
So far, eleven circumbinary planets have been discovered by \textit{Kepler}
with Kepler-453b and Kepler-1647b constituting number 10 and 11, as
reported by \cite{wel15} and \cite{kos16}, respectively.

The main purpose of the \textit{Kepler} mission is to identify
exoplanets via the transit method near or within the host star's
habitable zone (HZ).  The lion's share of stars-of-study encompass
main-sequence stars of spectral types G, K, and M, with latter ones
also referred to as red dwarfs.  Recent catalogs of stars studied by 
\textit{Kepler} have been given by \cite{kir16} and \cite{tho17}.
Here \cite{tho17} offer the latest results for the general catalog
from \textit{Kepler}, as it contains all observed objects, including
circumbinary planets, potentially habitable planets, and (most likely)
non-habitable planets.  On the other hand, the catalog by \cite{kir16}
is mostly focused on eclipsing binary systems.

Previous theoretical work about circumbinary planets in binary systems
has been given by, e.g., \cite{kan13}, \cite{egg13}, \cite{hag13},
\cite{cun14,cun15}, \cite{zul16}, \cite{pop17}, \cite{she17}, and
\cite{wan17}, and references therein.  These types of studies focus on
the formation, orbital stability, secular evolution, and/or environmental
forcings pertaining to those systems.  For example, recently, \cite{wan17}
presented fitting formulae for the quick determination of the existence
of P-type HZs in binary systems.  Objects hosted by P-type systems which
might be potentially habitable could include exoplanets, exomoons, and
exo-Trojans.  For Kepler-16, the latter two kinds of objects have been
discussed by Quarles et al. (2012), hereafter QMC12. 

Kepler-16(AB) is a pivotal example of a planet-hosting binary; it is
61 parsecs (199 light years) from Earth (see Table~1); for more
detailed information see \cite{doy11}, and references therein.
The system consists of the primary star, Kepler-16A, a K-dwarf
of about 0.69~$M_\odot$, and the secondary star, Kepler-16B, a
red dwarf star.  The circumbinary planet of that system
is similar to Saturn in mass and density.
Kepler-16b has a nearly circular orbit with
an eccentricity of approximately 0.007 and a small deviation
in orbital inclination to that of its host stars indicating
that it may have formed within the same circumbinary disk as
the two stars. Although Kepler-16b proves to be an interesting
exoplanet, it is considered to be cold, gaseous, and ultimately
uninhabitable.  However, previous work by QMC12 has focused on
the possibility of both Earth-mass exomoons and Trojans, which
if existing may be potentially habitable.  Among other
considerations, we intend to expand the work by QMC12 in this
article.

The structure of our paper is as follows.  In Section~2, we
report the stellar parameters.  A special effort is made to
determine the effective temperature of Kepler-16B.  Section~3
discusses the HZ of the Kepler-16(AB) binary system in consideration
of different types of climate models available in the literature.
For tutorial reasons, we also discuss the HZ of Kepler-16A, with
Kepler-16B assumed absent.  In Section~4, we consider the
previous results by QMC12 for Earth-mass moons and Trojans
in relationship to Kepler-16's HZ.  Furthermore, additional
stability simulations based on a modified version of the
mercury6 integration package are pursued to explore the
possible parameter space of stable objects in the Kepler-16(AB)
system.  Our summary and conclusions are given in Section~5.

%%%%%%%%%%%%%%%%%%%%%%%%%%%%%%%%%%%%%%%%%%%%%%%%%%%%%%%%%%%%%%%%%%%%%%%%

\section{Stellar Parameters}

Regarding our study, stellar parameters are of pivotal importance for
the calculation of stellar HZs as well as for orbital stability simulations
of possible exomoons and Trojan objects.  Most relevant parameters of the
Kepler-16(AB) system have been previously reported by \cite{doy11}, 
as they announce a transiting circumbinary planet observed by the
\textit{Kepler} spacecraft.  Kepler-16A was identified as a K-type
main-sequence star with effective temperature, radius, and mass given
as (see Table~1) $4450 \pm 150$~K, $0.6489 \pm 0.0013$~$R_\odot$, and
$0.6897 \pm 0.0035$~$M_\odot$, respectively.  Here the relative
uncertainty bar is largest for the stellar effective temperature (see
Table~2).

However, less information has been conveyed for Kepler-16B, which based on
its mass of about 0.20255~$M_\odot$ \citep{doy11} is identified as a
red dwarf.  But Kepler-16B's effective temperature needs to be determined
as well to compute the HZ for the Kepler-16 binary system.  Thus, to
determine Kepler-16B's stellar effective temperature, we utilize the mass
-- effective temperature relationship by \cite{man13}.  They have
analyzed moderate resolution spectra for a set of nearby K and M dwarfs
with well-known parallaxes and interferometrically determined radii to define
their effective temperatures, among other quantities.  They also adopt
state-of-the-art PHOENIX atmosphere models, as described.  Thus, we conclude
that the effective temperature of Kepler-16B is $3308 \pm 110$~K (see
Fig.~1).  Here the uncertainty bar has been estimated based on the results
of similar objects included in the sample.  From other work as, e.g.,
that by \cite{kir91} and \cite{bar98} the spectral type of Kepler-16B has
been deduced as $\sim$M3~V.  Both the effective temperature and radius
of Kepler-16B are important for determining the different types of HZs
of the Kepler-16(AB) system (see Sect.~4).

%%%%%%%%%%%%%%%%%%%%%%%%%%%%%%%%%%%%%%%%%%%%%%%%%%%%%%%%%%%%%%%%%%%%%%%%

\section{The Kepler-16 Habitable Zone}

A crucial aspect of this study is the evaluation of Kepler-16's HZ.  
The HZ is a region around a star or a system of stars in which
terrestrial planets could potentially have surface temperatures at
which liquid water could exist, given a sufficiently dense atmosphere
\cite[e.g.,][]{kas93,jon01,und03}.  When determining the HZ, both
inner limits and outer limits are calculated, in response to different
types of criteria, thus defining the HZ.  The determination of the
location of the HZ is significant in the context of theoretical
studies as well as for the purpose of planet search missions
\cite[e.g.,][and references therein]{lam09,kas14,kal17}.

Inner limits previously used for stellar HZs include those set
by the recent Venus (RV), the runaway greenhouse effect, and the
onset of water loss.  Furthermore, outer limit of the stellar HZ
has been set by the first CO$_2$ condensation, the maximum greenhouse
effect for a cloud-free CO$_2$ atmosphere and the early Mars (EM) setting.
For example, \cite{kas93} describe the runaway greenhouse effect such
that the greenhouse phenomenon is enhanced by water vapor, thus
promoting surface warming.  The latter further increases the atmospheric
vapor content, thus resulting in an additional rise of the planet's
surface temperature.  Consequently, this will lead to the rapid evaporation
of all surface water.  On the other hand \citep[see, e.g.,][]{und03}, the
water loss criterion means that an atmosphere is warm enough to have
a wet stratosphere, from where water is gradually lost by atmospheric
chemical processes to space.

Table~3 shows the HZ limits for Kepler-16A, treated as a single star,
for tutorial reasons.  Here GHZ denotes the general habitable zone,
bracketed by the runaway greenhouse and maximum greenhouse criteria,
whereas RVEM denotes the kind of HZ, defined by the settings of recent
Venus and early Mars; this latter type of HZ is also sometimes referred
to as (most) optimistic HZ; see, e.g., \cite{kal17} and references
therein.  Figure~2 and Tables~3 and 4 convey the results for the various
HZ limits as well as for the GHZ and RVEM.  The most recent results
based on \cite{kop13,kop14} have been included as well, which indicate
updated HZ limits.  For the inner and outer HZ limits, they assumed
$\mathrm{H}_2\mathrm{O}$ and $\mathrm{CO}_2$ dominated atmospheres,
respectively, while scaling the background $\mathrm{N}_2$ atmospheric
pressure with the radius of the planet. Moreover, from said climate
model, several equations were generated, which correspond to select
inner and outer HZ limit criteria.

Surely, most of our study focuses on Kepler-16 as a binary thus taking
into account both Kepler-16A (an orange dwarf) and Kepler-16B (a red
dwarf); see Table~1 for data.  The computation of the GHZ and RVEM of
Kepler-16(AB) follows the work by \cite{cun14,cun15} and \cite{wan17}.
Information is given in Fig.~3; here RHZ refers to the so-called
radiative habitable zone (applicable to both the GHZ and RVEM),
which is based on the planetary climate enforcements
set by both stellar components, while deliberately ignoring
the orbital stability criterion regarding a possible system planet.
Figure~3 indicates the inner and outer RHZ limits with the inner HZ
limit defined as the maximum radial distance of the inner RHZ (red lines)
and the outer HZ limit defined as the minimum radial distance of the
outer RHZ (blue lines).  This approach conveys the HZ region for GHZ
(darkest green) and RVEM (medium green) criteria (see also Table~5).
As expected the RVEM criteria produces a more generous HZ region.
We also indicate the orbital stability limit (black dashed line)
based on \cite{hol99}, referred to as $a_{\rm orb}$.  In fact it
is found that the widths of the GHZ and RVEM for Kepler-16(AB)
are significantly less than for Kepler-16A (single star approach),
owing to the additional criterion of orbital stability for possible
system planets.

Previous work by \cite{mis00} argues that the HZ about a main-sequence
star might be further extended if $\mathrm{CO}_2$ cloud coverage is
assumed.  In the case of the Sun, this assumption would amount to an
outer limit of 2.40~AU for the hereupon defined extended habitable
zone (EHZ)\footnote{The previous work by \cite{mis00} has been superseded
by more recent studies, including those given by \cite{hal09}, \cite{pie11},
\cite{wor13},
and \cite{kit16}; see also summary by \cite{sea13}.  For example,
\cite{kit16} argued that the heating assumed by \cite{mis00} has
been overestimated, thus putting the extension of the outer HZ in
question.  However, in the following, we will parameterize the outer
limit of \cite{mis00}, and the significance of our results will not
rely on the full extent of the HZ introduced by them.  Moreover,
\cite{pie11} argued that planetary HZs could extend to up to 10~AU
for single G-type stars (or, say, about 3~AU for single K-type stars,
as indicated by their Fig.~1), which is well beyond
the outer limit advocated by \cite{mis00}.}.
\cite{blo07} have explored the habitability around Gliese~581
with focus on the possible planet GJ~581d.  They argue that the RHZ
could be further extended if the atmospheric structure is determined
by particularly high base pressures.  Thus, the outer limit for the
EHZ is not very well constrained, but could be parameterized as
$\epsilon \sqrt{L}$ with $\epsilon$ in the likely range between 2.0 and
3.0 and $L$ defined as stellar luminosity (in units of solar luminosity).
Hence, $\epsilon = 2.4$ corresponds to the value of \cite{mis00}.
Results for the EHZ of Kepler-16(AB) are given in Figure~4 and Table~6.

Another aspect of study is that concerning the GHZ and RVEM, we also
have explored the impact of the observational uncertainties of the
stellar luminosities on inner and outer limits of the RHZs (see
Figs.~5 and 6).  It is found that the uncertainty in the stellar
luminosity ${\Delta}L$ moves the inner and outer limits of both the
GHZ and the RVEM by about $\pm$6\%.  Our results are summarized in Table~7.
Here we also see that the inner limits of both the GHZ and RVEM are set
by the additional criterion of orbital stability regarding possible
circumbinary planets, referred to by \cite{cun14} as PT habitability.
Additionally, it is found that the HZ around Kepler-16A (if treated as
a single star) would be significantly more extended than the HZ of
Kepler-16(AB).  Thus, Kepler-16B notably reduces the prospect of
habitability in that system.

%%%%%%%%%%%%%%%%%%%%%%%%%%%%%%%%%%%%%%%%%%%%%%%%%%%%%%%%%%%%%%%%%%%%%%%%

\section{Stability Investigations for Earth-Mass Exomoons and Trojans}

Previously, QMC12 have exemplary case studies for the orbital stability of 
Earth-mass objects (i.e., Trojan exoplanet or exomoon) in the Kepler-16(AB) system.
Their numerical methods were based on the Wisdom-Holman mapping technique
and the Gragg-Burlisch-Stoer algorithm \citep{gra96}.  The resulting
equations of motion were integrated forward in time for 1 million years
using a fixed/initial (WH/GBS) time step.  QMC12 showed that, in principle,
both Trojan exoplanets and exomoons are able to exist in the Kepler-16(AB)
system.  Figures 7 and 8 show the results by QMC12 together with the
updated system's HZs, i.e., the GHZ, RVEM, and EHZ.  It is found that
the orbital settings of those objects are consistent with an EHZ
(with $\epsilon \lta 2.2$) or with the RVEM if upper limits of the
stellar luminosities, consistent with the observational uncertainties,
are considered.

In order to better understand the dynamical domain of possible exo-Trojans,
we perform additional 5,000 stability simulations using a modified version
of the mercury6 integration package that is optimized for circumbinary systems
\citep{cha02}.  In these simulations, we adopt the orbital parameters from
\cite{doy11} for the binary components and the Saturnian planet.  We also
consider Earth-mass objects with different initial conditions.
Table~8 conveys the initial conditions for exomoon sample cases, which are:
the semimajor axis $a$, eccentricity $e$, inclination $i$, argument of periastron
$\omega$, and mean anomaly $M$ for each body.  A simulation is terminated when
the Earth-mass body either crosses the binary orbit or has a radial distance
from the center of mass greater than 100~AU; this will be viewed as an ejection.

The orbital evolution of the four bodies are evaluated on a 10 Myr timescale.
The initial orbital elements are chosen using uniform distributions.  The initial
semimajor axis of the Earth-mass object is selected from values ranging from
0.6875~AU to 0.7221~AU (i.e., $\pm$ 0.5 Hill radii); furthermore, eccentricities
are limited to 0.1 and inclinations are limited to 1$^\circ$.  The initial argument
of periastron and mean anomalies are selected randomly between 0$^\circ$ and
360$^\circ$.  The statistical distributions of the surviving population are shown
in Figure~9 to illustrate possible correlations between parameters.

Overall $\sim$10\% of the simulations (496) are identified as stable
(i.e., survived for 10 Myr) as depicted in Figure 10.  By delineating
the stable (cyan) and unstable (gray) points, it is seen that the stable
initial conditions correspond to Trojans and are separated in relative phase
from Kepler-16b by $\sim$60$^\circ$ to 90$^\circ$.  This also appears in
Figure 9 through the distribution for $\lambda^*$, the relative mean longitude.
The inclinations of the orbitally stable Earth-mass objects in Fig. 9 remain
uniformly distributed and thus are unlikely to affect the overall stability.
Figure 11 illustrates the orbital evolution in a rotated-reference frame
of two initial conditions taken from Figure 10.  The panels of Fig. 11 show
the first $\sim$1,000 years of orbital evolution, where the run in the
top panel would continue in a Trojan orbit for the 10 Myr simulation time
and the other run (bottom panel) evolves in a horseshoe orbit, which quickly
becomes unstable.  We also found that the eccentricity distribution as
obtained prefers values close to circular, whereas the relative mean
longitude distribution reflects, by a factor of two, more trailing orbits
than preceding orbits.

%%%%%%%%%%%%%%%%%%%%%%%%%%%%%%%%%%%%%%%%%%%%%%%%%%%%%%%%%%%%%%%%%%%%%%%%

\section{Summary and Conclusions}

The purpose of our study is to continue investigating the habitable zone
as well as the general possibility of Earth-mass exomoons and Trojans in
Kepler-16.  The binary system Kepler-16(AB) consists of a low-mass main-sequence
star, a red dwarf and a circumbinary Saturnian planet.  The temperatures of
the two stellar components are given as $4450 \pm 150$~K and $3308 \pm 110$~K,
respectively.  Previously, QMC12 pursued an exploratory study about this
system, indicating that based on orbital stability considerations both
Earth-mass exomoons and Earth-mass Trojan planets might be possible.
The aim of the present study is to offer a more thorough analysis of
this system.  We found the following results:

\bigskip
\noindent
(1) As previously said by QMC12, Kepler-16 possesses a circumbinary HZ;
its width depends on the adopted climate model.
Customarily, these HZs are referred to as GHZ and RVEM; the latter is also
sometimes referred to as optimistic HZ \citep[e.g.,][]{kop13, kal17}.  For
objects of thick CO$_2$ atmospheres including clouds, the HZ is assumed to
be further extended, thus giving rise to the EHZ as proposed by \cite{mis00}.

\bigskip
\noindent
(2) Our work confirms earlier simulations by QMC12 that both Earth-mass
exomoons and Earth-mass Trojan planets could stably orbit in
that system.  However, in this study, we adopted longer timescales and
also explored the distributions of eccentricity and inclinations of the
Earth-mass test objects considered in our study. 

\bigskip
\noindent
(3) Exomoons and Trojans, associated with the Saturnian planet, are found
to be situated in the lower portion of the EHZ (i.e., $\epsilon \lta 2.2$).
A more detailed analysis also implies that the distances of those objects
may be consistent with the RVEM (i.e., optimistic HZ) if a relatively
high luminosity for the stellar components is assumed (but still consistent
with the uncertainty bars) or if the objects are allowed to temporarily leave
the RVEM-HZ without losing habitability.  The latter property is maintained
if habitability is provided by a relatively thick atmosphere \cite{wil02}.

\bigskip
\noindent
(4) For tutorial reasons, we also compared the HZ of the system's primary to
that of the binary system.  We found that the latter is reduced by 42\% (GHZ)
and 48\% (RVEM) despite the system's increase in total luminosity given by
the M-dwarf.  The reason is that for the binary, the RHZ is unbalanced and
it is further reduced by the additional requirement of orbital stability as
pointed out previously \citep[e.g.,][]{egg13,cun14}.

\bigskip
\noindent
(5) Moreover, we pursued new stability simulations for Earth-mass objects
while taking into account more general initial conditions.  The attained
eccentricity distribution prefers values close to circular, whereas the
inclination distribution is relatively flat.  The distribution in the
initial relative phase indicates that the stable solutions are distributed
near the co-orbital Lagrangian points, thus increasing the plausibility
for the existence of those objects.

Our study shows that the binary system Kepler-16(AB) has a HZ of notable extent,
though smaller than implied by the single-star approach, with its extent
critically depending on the assumed climate model for the possible Earth-mass
Trojan planet or exomoon.  Thus, Kepler-16 should be considered a valuable
target for future planetary search missions.  Moreover, it is understood that
comprehensive studies of habitability should take into account additional
forcings by planet host stars, such as stellar activity and strong winds
expected to impact planetary conditions as indicated through analyses by,
e.g., \cite{lam07}, \cite{tar07}, \cite{lam09}, \cite{kas14}, and \cite{kal17}.
Recent articles about the impact on stellar activity on prebiotic environmental
conditions have been given by, e.g., \cite{cun16} and \cite{air17}.

%%%%%%%%%%%%%%%%%%%%%%%%%%%%%%%%%%%%%%%%%%%%%%%%%%%%%%%%%%%%%%%%%%%%%%%%%%%%%%

\acknowledgments
This work has been supported by the Department of Physics, University
of Texas at Arlington (UTA).  The simulations presented here were performed
using the OU Supercomputing Center for Education \& Research (OSCER) at the
University of Oklahoma (OU).  Furthermore, we would like to thank the
anonymous referee for useful suggestions, allowing us to improve the manuscript.

%%%%%%%%%%%%%%%%%%%%%%%%%%%%%%%%%%%%%%%%%%%%%%%%%%%%%%%%%%%%%%%%%%%%%%%%%%%%%%

\clearpage

%\bibliographystyle{natbib}

%%%%%%%%%%%%%%%%%%%%%%%%%%%%%%%%%%%%%%%%%%%%%%%%%%%%%%%%%%%%%%%%%%%%%%%%

%%
%%  FIGURES
%%

%+++++++++++++++++++++++++++++++++++++++++++++++++++++++++++++++++++++++

\clearpage

%%% *** Fig.1
%%%%%%%%%%%%%%%%%%%%%%%%%%%%%%%%%%%%%%%%%%%%%%%%%%%%%%%%%%%%%%%%%
\begin{figure*} 
\centering
\begin{tabular}{c}
\epsfig{file=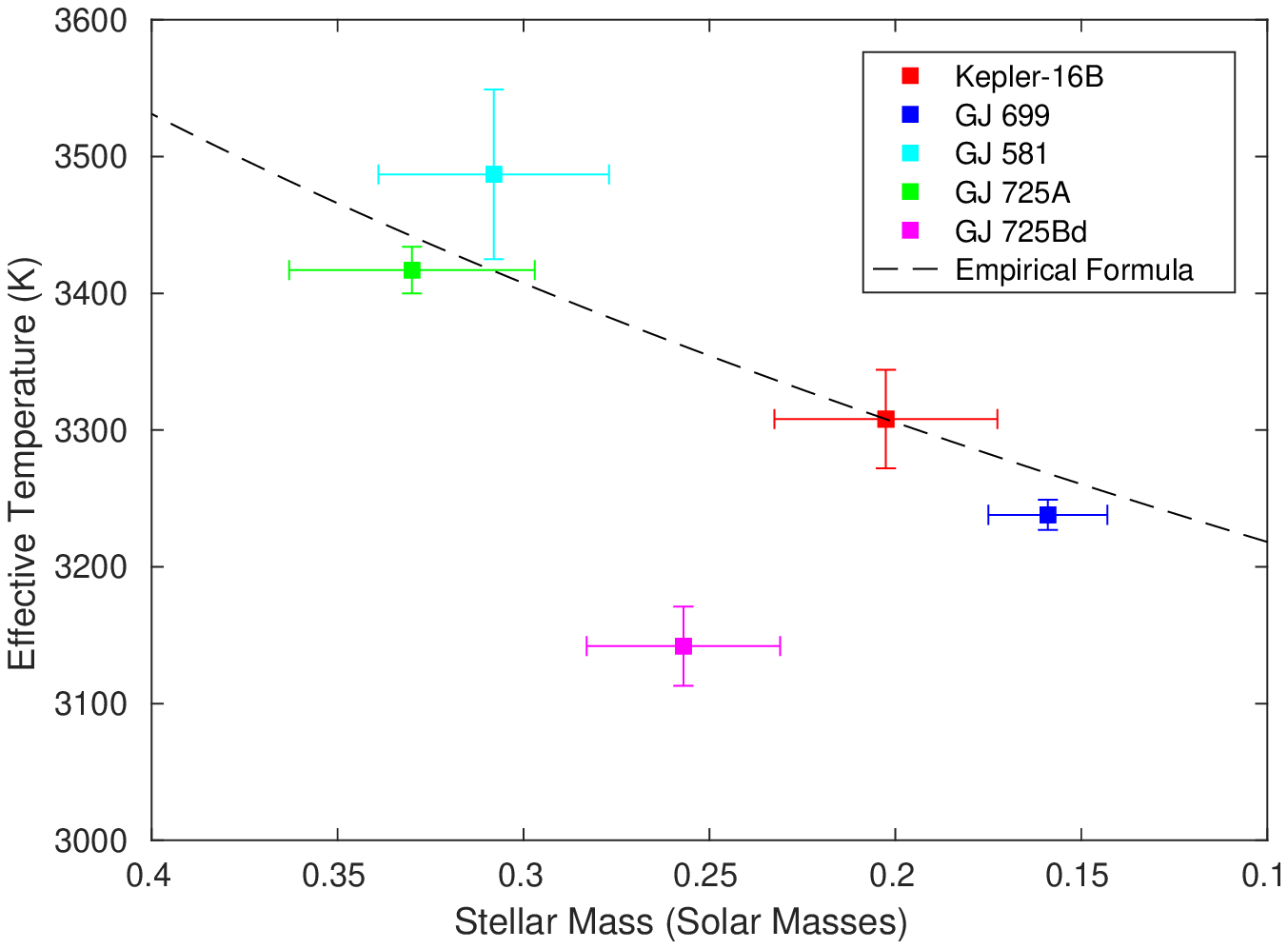,width=0.85\linewidth}
\end{tabular}
\caption{
Depiction of the effective temperature of Kepler-16B determined via an
empirical formula given by \cite{man13} that relates the mass to the
effective temperature, and vice versa, for M dwarf stars.  By knowing
the mass of Kepler-16B, its effective temperature can be extracted,
resulting in an effective temperature of $3308 \pm 110$~K.  In addition,
a subset of the sample of M dwarf stars is depicted, used to derive
the adopted empirical formula.
}
\end{figure*}

%+++++++++++++++++++++++++++++++++++++++++++++++++++++++++++++++++++++++

\clearpage

%%% *** Fig.2
%%%%%%%%%%%%%%%%%%%%%%%%%%%%%%%%%%%%%%%%%%%%%%%%%%%%%%%%%%%%%%%%%
\begin{figure*} 
\centering
\begin{tabular}{c}
\epsfig{file=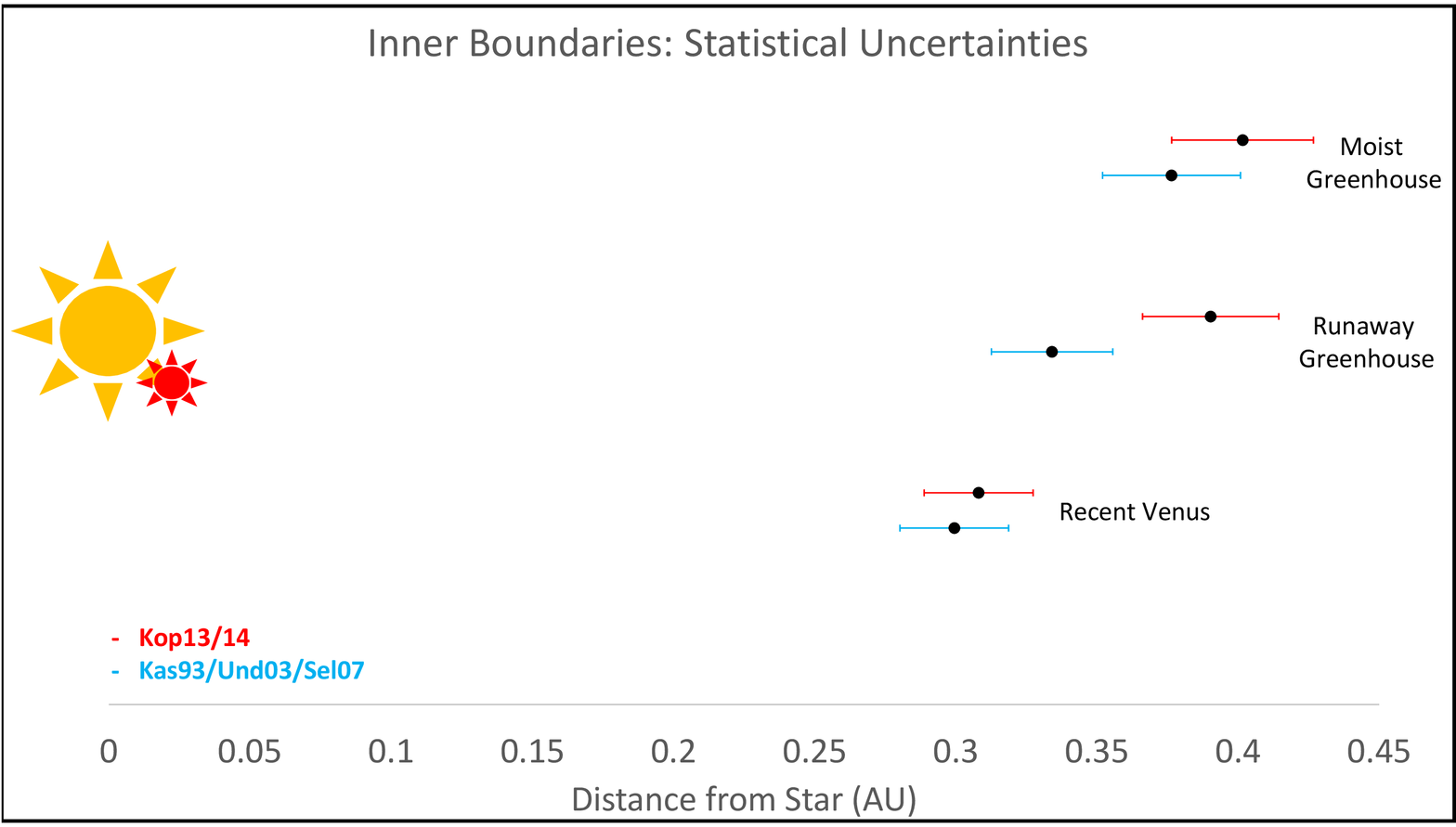,angle=270,width=0.65\linewidth} \\
\epsfig{file=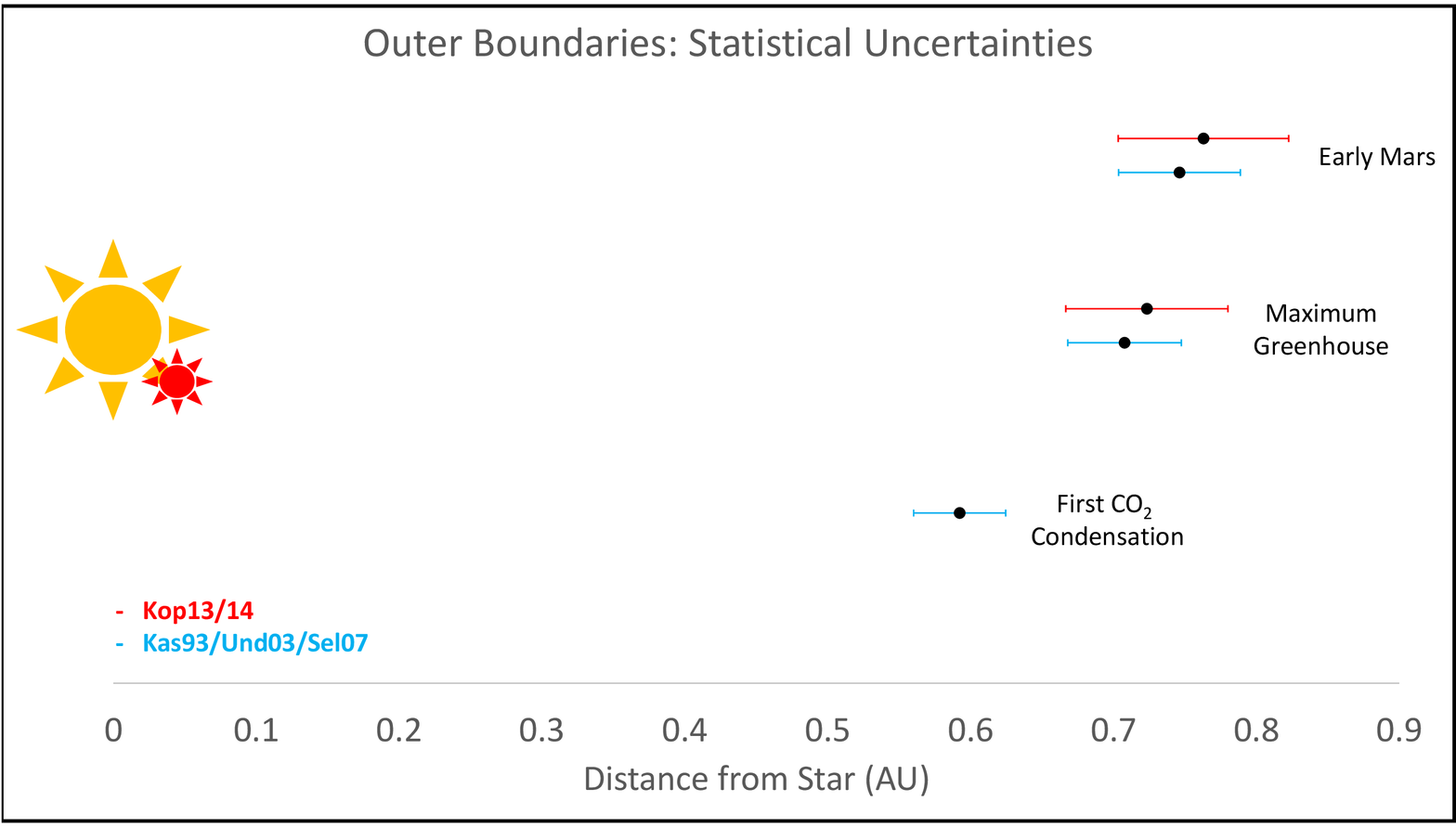,angle=270,width=0.65\linewidth}
\end{tabular}
\caption{
Inner and outer HZ limits for Kepler-16A (single star approach) while comparing
two different determination methods.  We also include information on the
inherent statistical uncertainties of those based on \cite{pre86} (see also
Table~3).  The blue data correspond to the inner and outer HZ boundaries as
expected from utilizing the method of \cite{kas93} with updates by \cite{und03}
and \cite{sel07}.  Conversely, the red data correspond to the inner and outer
HZ limits as expected from utilizing the method specified by \cite{kop13,kop14}. 
}
\end{figure*}

%+++++++++++++++++++++++++++++++++++++++++++++++++++++++++++++++++++++++

\clearpage

%%% *** Fig.3
%%%%%%%%%%%%%%%%%%%%%%%%%%%%%%%%%%%%%%%%%%%%%%%%%%%%%%%%%%%%%%%%%
\begin{figure*} 
\centering
\begin{tabular}{c}
\epsfig{file=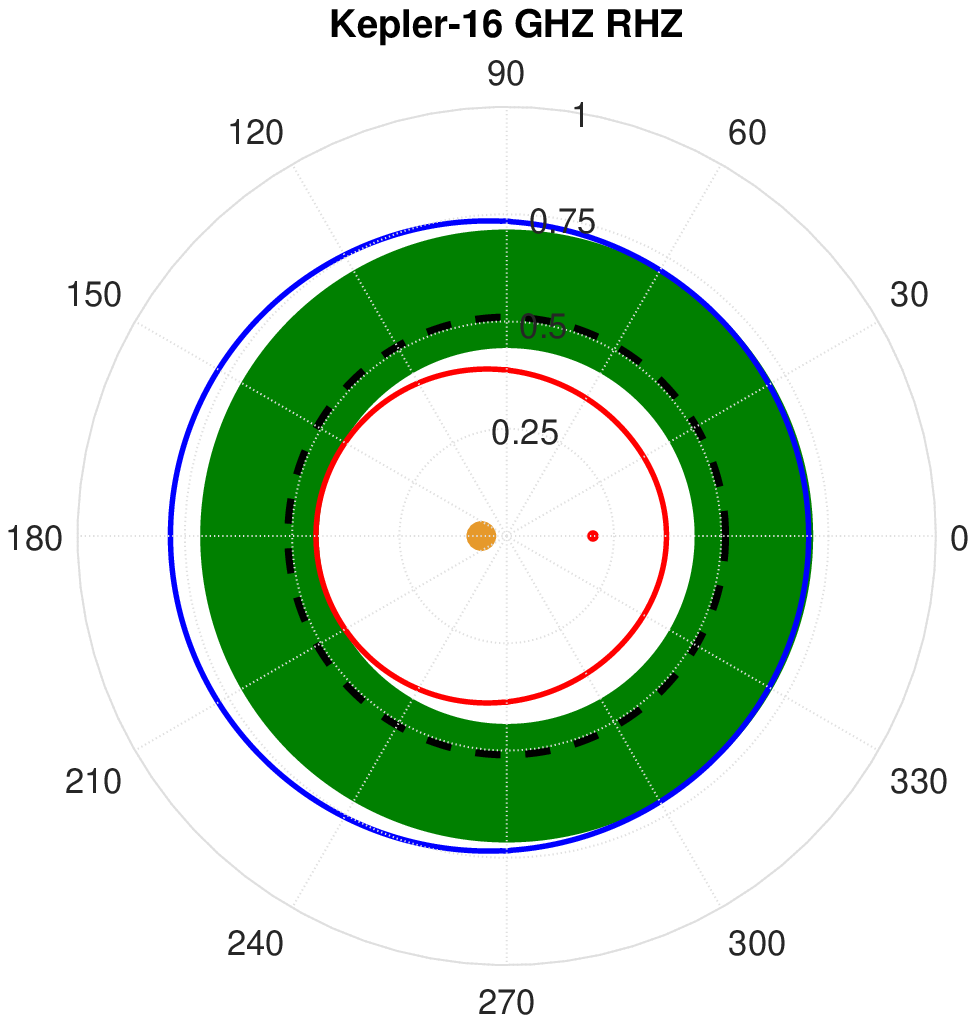,width=0.45\linewidth} \\
\noalign{\bigskip}
\noalign{\bigskip}
\epsfig{file=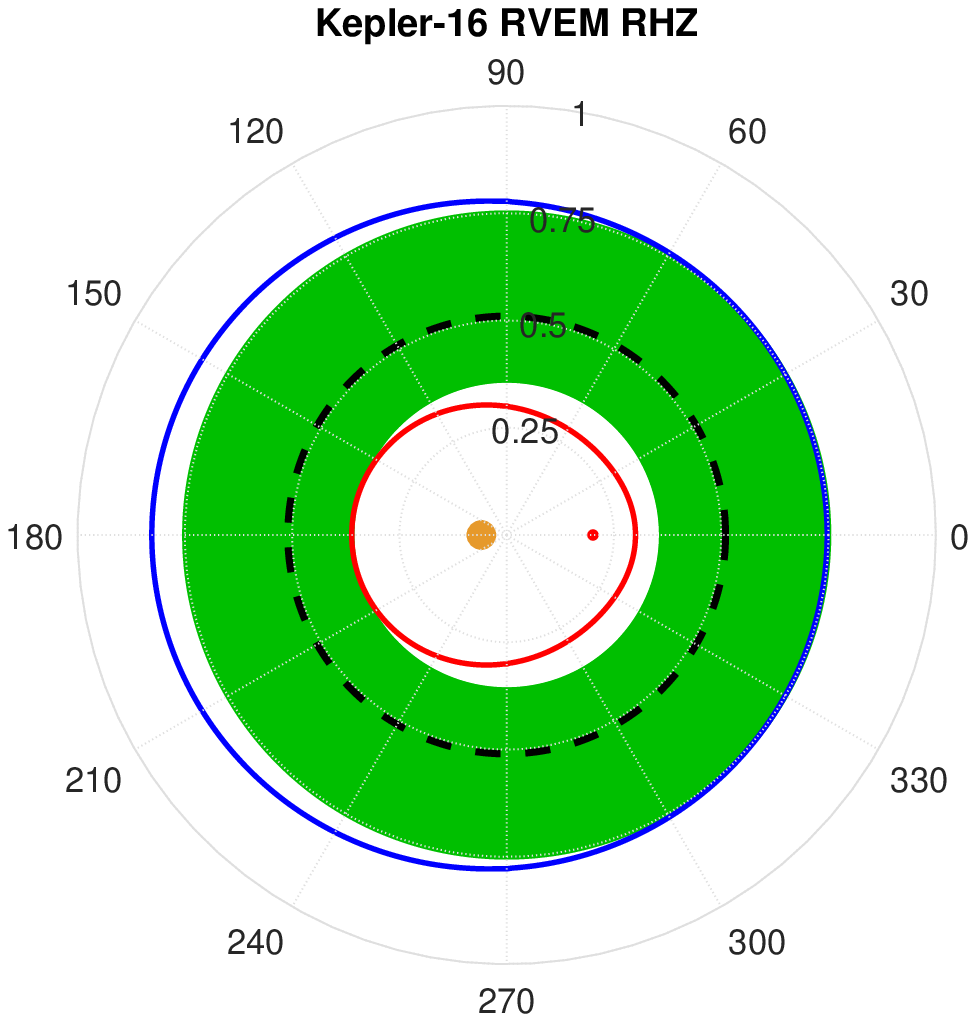,width=0.45\linewidth}
\end{tabular}
\caption{
Depiction of the RHZ for the GHZ and RVEM criteria based on methods given by
\cite{cun14,cun15} and \cite{wan17}.  In both plots the red and blue lines correspond
to the inner and outer RHZ limits with the inner HZ limit defined as the maximum radial
distance of the inner RHZ (red lines) and the outer HZ limit defined as the minimum
radial distance of the outer RHZ (blue lines).  This approach produces the conventional
HZ region for GHZ (darkest green) and RVEM (medium green) criteria.  As expected the
RVEM criteria produces a more generous HZ region as shown.  Lastly, the black dashed line
represents the orbital stability limit, calculated using the formula provided by \cite{hol99}
for P-type orbits, in which bodies exterior to that line are orbitally stable while bodies
interior to that line are orbitally unstable.
}
\end{figure*}

%+++++++++++++++++++++++++++++++++++++++++++++++++++++++++++++++++++++++

\clearpage

%%% *** Fig.4
%%%%%%%%%%%%%%%%%%%%%%%%%%%%%%%%%%%%%%%%%%%%%%%%%%%%%%%%%%%%%%%%%
\begin{figure*} 
\centering
\begin{tabular}{c}
\epsfig{file=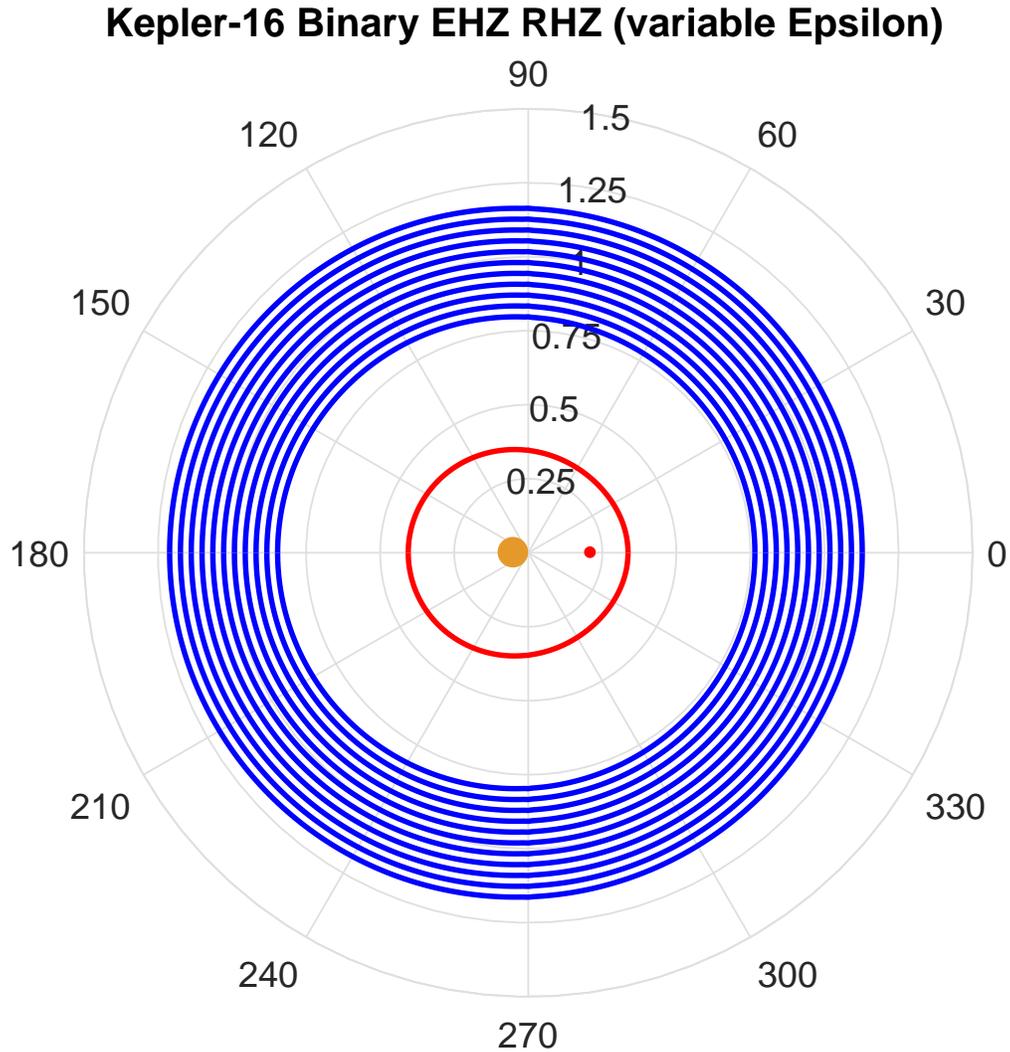,width=0.80\linewidth}
\end{tabular}
\caption{
Different outer boundaries (blue lines) of the EHZ resulting from the
different epsilon values ranging from $\epsilon=2.0$ (innermost blue line)
to $\epsilon=3.0$ (outermost blue line).  A median value of $\epsilon=2.5$
has been chosen for our definition of the EHZ akin to \cite{mis00}, which
is also adopted for our analysis in the subsequent Figs.~7, 8, and 10 and
depicted as the lightest green regions. 
}
\end{figure*}

%+++++++++++++++++++++++++++++++++++++++++++++++++++++++++++++++++++++++

\clearpage

%%% *** Fig.5
%%%%%%%%%%%%%%%%%%%%%%%%%%%%%%%%%%%%%%%%%%%%%%%%%%%%%%%%%%%%%%%%%
\begin{figure*} 
\centering
\begin{tabular}{c}
\epsfig{file=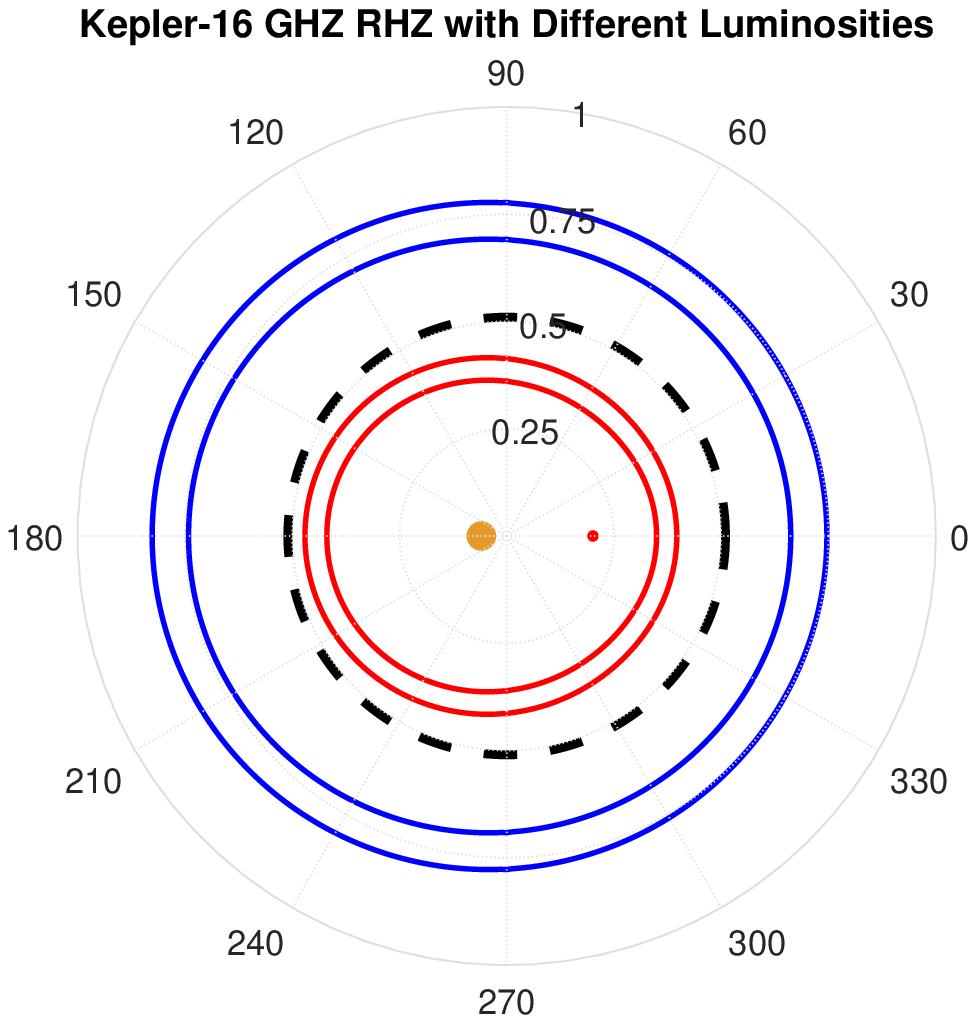,width=0.45\linewidth} \\
\noalign{\bigskip}
\noalign{\bigskip}
\epsfig{file=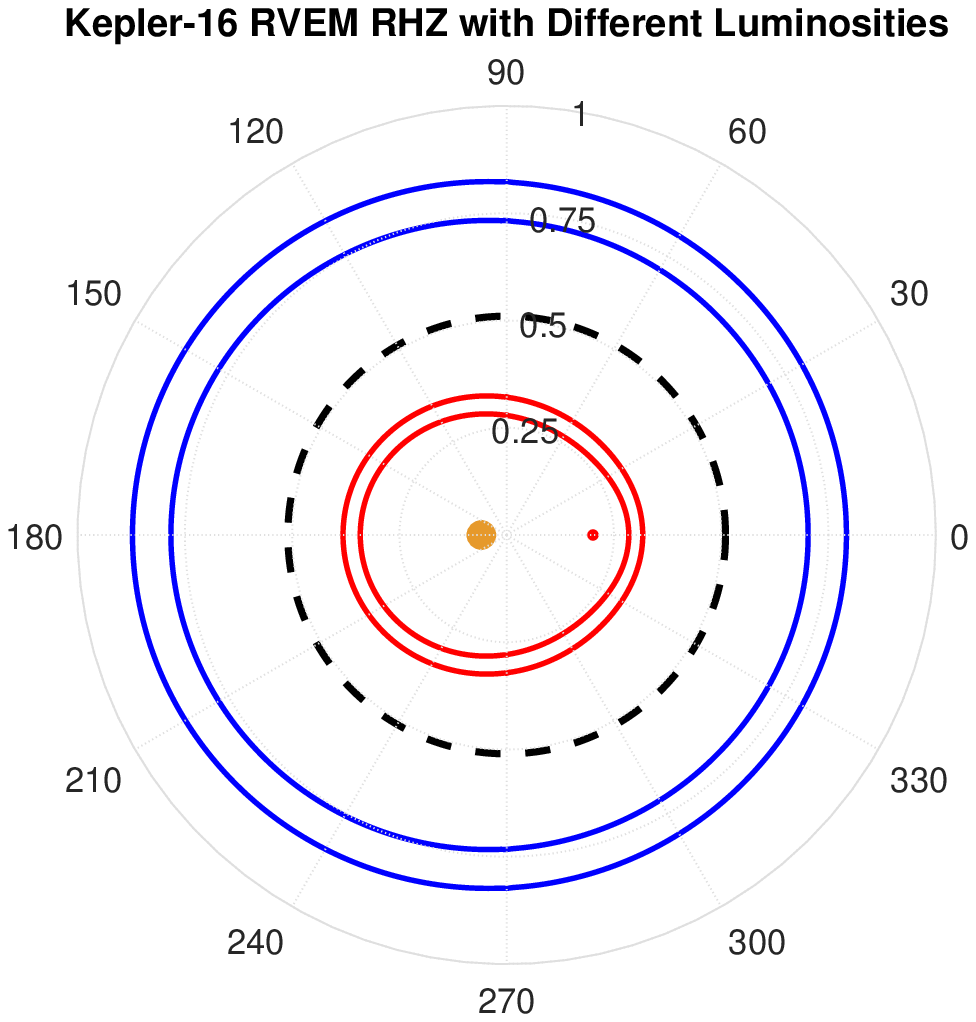,width=0.45\linewidth}
\end{tabular}
\caption{
Depiction of the inner and outer boundaries of the GHZ and RVEM, while utilizing
the upper and lower bounds of the stellar luminosities to illustrate how the
uncertainty in the luminosity affects the determination of the HZs.  In both
plots the inner and outer HZ limits are shown in red and blue, respectively,
with the inner sets of red and blue lines corresponding to the lower bound
luminosity and the outer sets of red and blue lines corresponding to the upper
bound luminosity.  As expected, the upper bound luminosity shifts the GHZ and
RVEM limits outward while the lower bound luminosity shifts those limits inward.}
\end{figure*}

%+++++++++++++++++++++++++++++++++++++++++++++++++++++++++++++++++++++++

\clearpage

%%% *** Fig.6
%%%%%%%%%%%%%%%%%%%%%%%%%%%%%%%%%%%%%%%%%%%%%%%%%%%%%%%%%%%%%%%%%
\begin{figure*} 
\centering
\begin{tabular}{c}
\epsfig{file=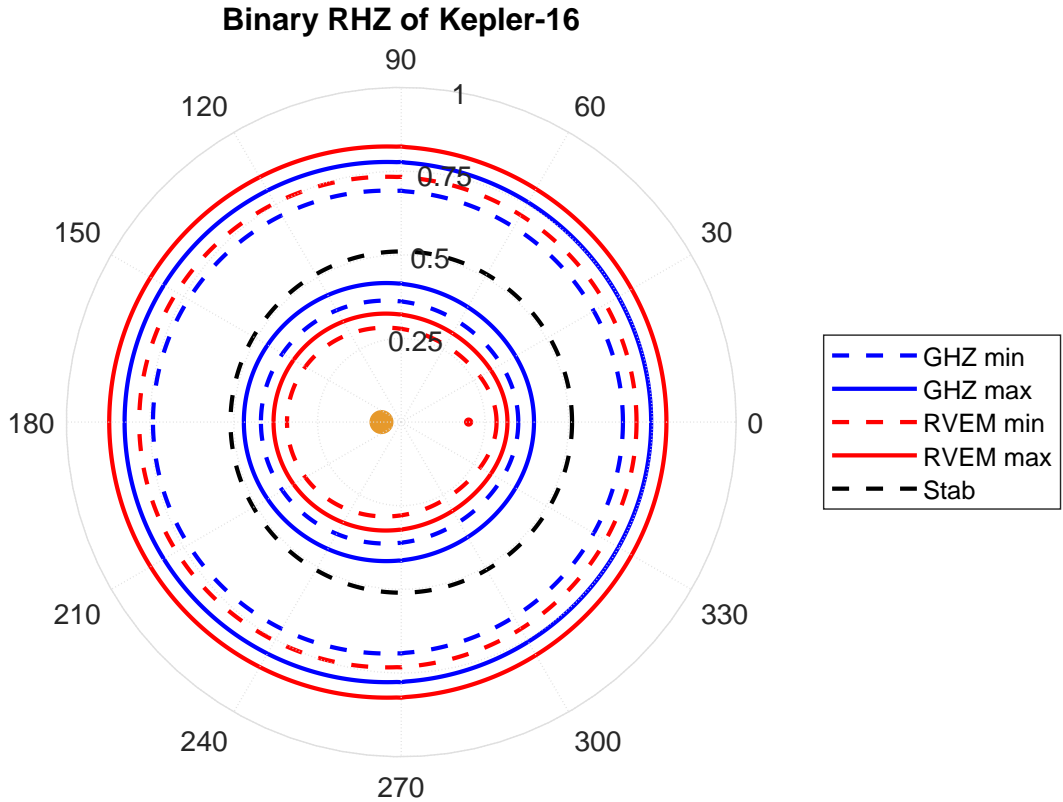,width=0.85\linewidth}
\end{tabular}
\caption{
Similar to Fig. 5, as this figure combines the inner and outer boundaries
for the GHZ and RVEM criteria while also incorporating the upper and lower
luminosity bounds; its emphasis is to illustrate the extents of the
achievable HZs based on luminosity and HZ criteria specification.
Additionally, the black dashed line represents the orbital stability limit.
The blue and red dotted lines correspond to the minimum possible inner
limits (associated with the lower bound luminosity) for the GHZ and RVEM,
respectively. The blue and red solid lines correspond to the maximum
possible outer limits (associated with the upper bound luminosity)
for GHZ and RVEM, respectively.  
}
\end{figure*}

%+++++++++++++++++++++++++++++++++++++++++++++++++++++++++++++++++++++++

\clearpage

%%% *** Fig.7
%%%%%%%%%%%%%%%%%%%%%%%%%%%%%%%%%%%%%%%%%%%%%%%%%%%%%%%%%%%%%%%%%
\begin{figure*} 
\centering
\begin{tabular}{c}
\epsfig{file=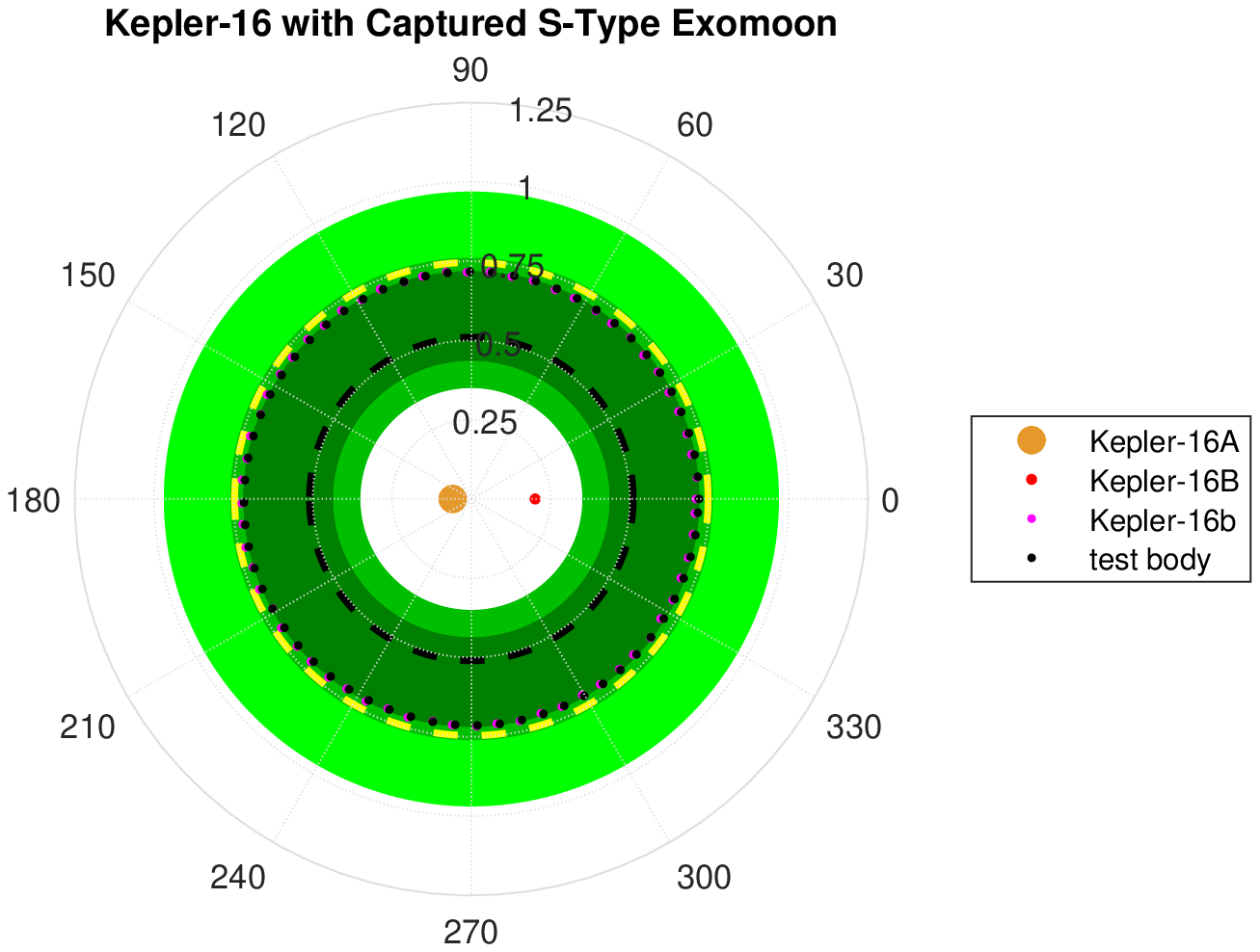,width=0.65\linewidth} \\
\noalign{\bigskip}
\noalign{\bigskip}
\epsfig{file=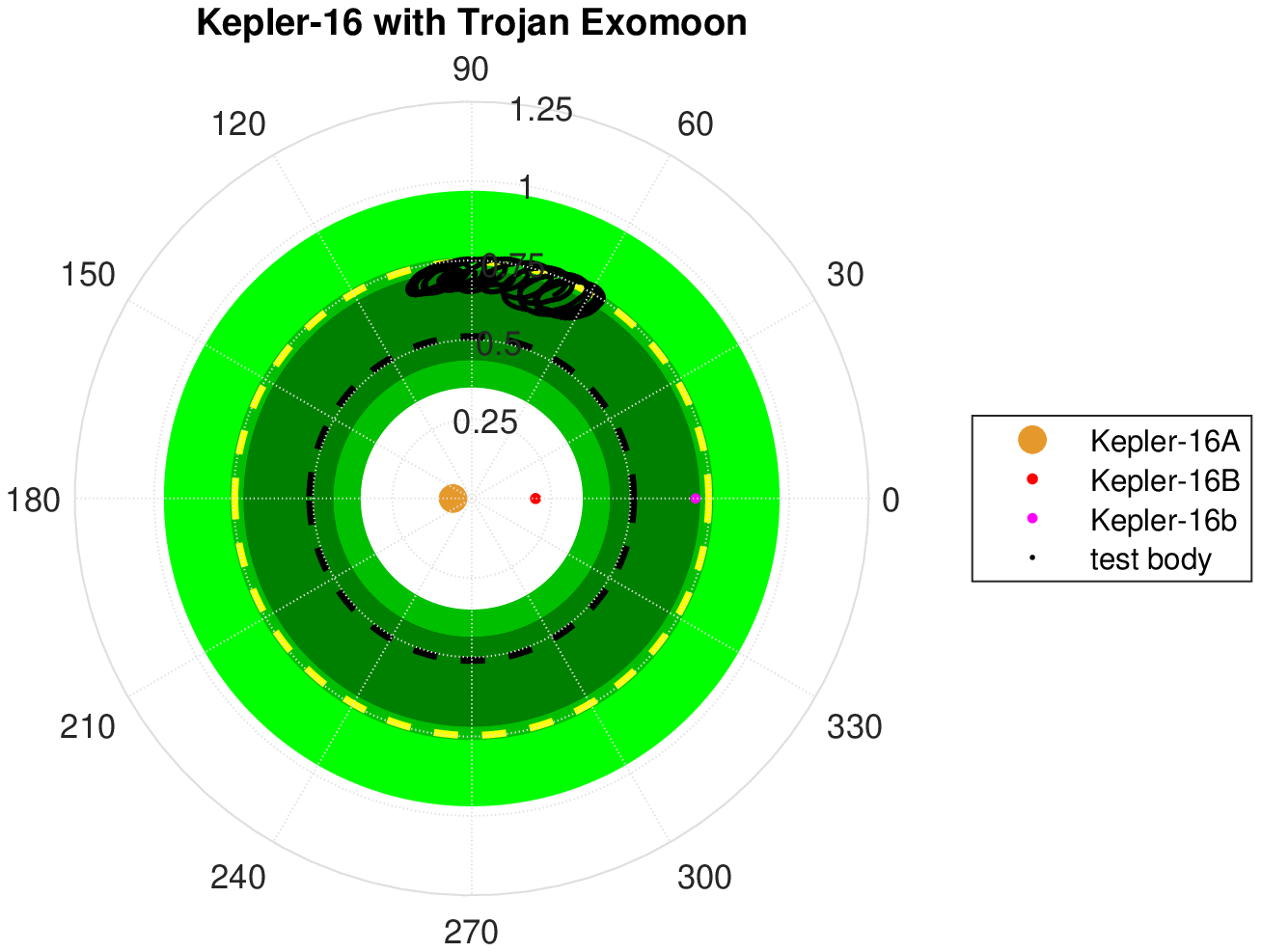,width=0.65\linewidth}
\end{tabular}
\caption{
Illustration of previous results by QMC12 with updated HZ regions.
(a) Depiction of an S-type captured Earth-mass exomoon
(black in QMC12); the primary and secondary stars (orange and red in QMC12,
respectively) and the Saturnian planet Kepler-16b are also given
(magenta in QMC12).  (b) Depiction of a possible Trojan exomoon in
a rotating reference frame (black in QMC12).  The darkest green
region represents the GHZ, the medium green region represents the RVEM,
and the lightest green region represents the EHZ.  The dashed yellow line
represents the outer limit of the GHZ if the stellar luminosities are
assumed at their upper limits as informed by the observational uncertainties.
}
\end{figure*}

%+++++++++++++++++++++++++++++++++++++++++++++++++++++++++++++++++++++++

%%% *** Fig.8
%%%%%%%%%%%%%%%%%%%%%%%%%%%%%%%%%%%%%%%%%%%%%%%%%%%%%%%%%%%%%%%%%
\begin{figure*} 
\centering
\begin{tabular}{c}
\epsfig{file=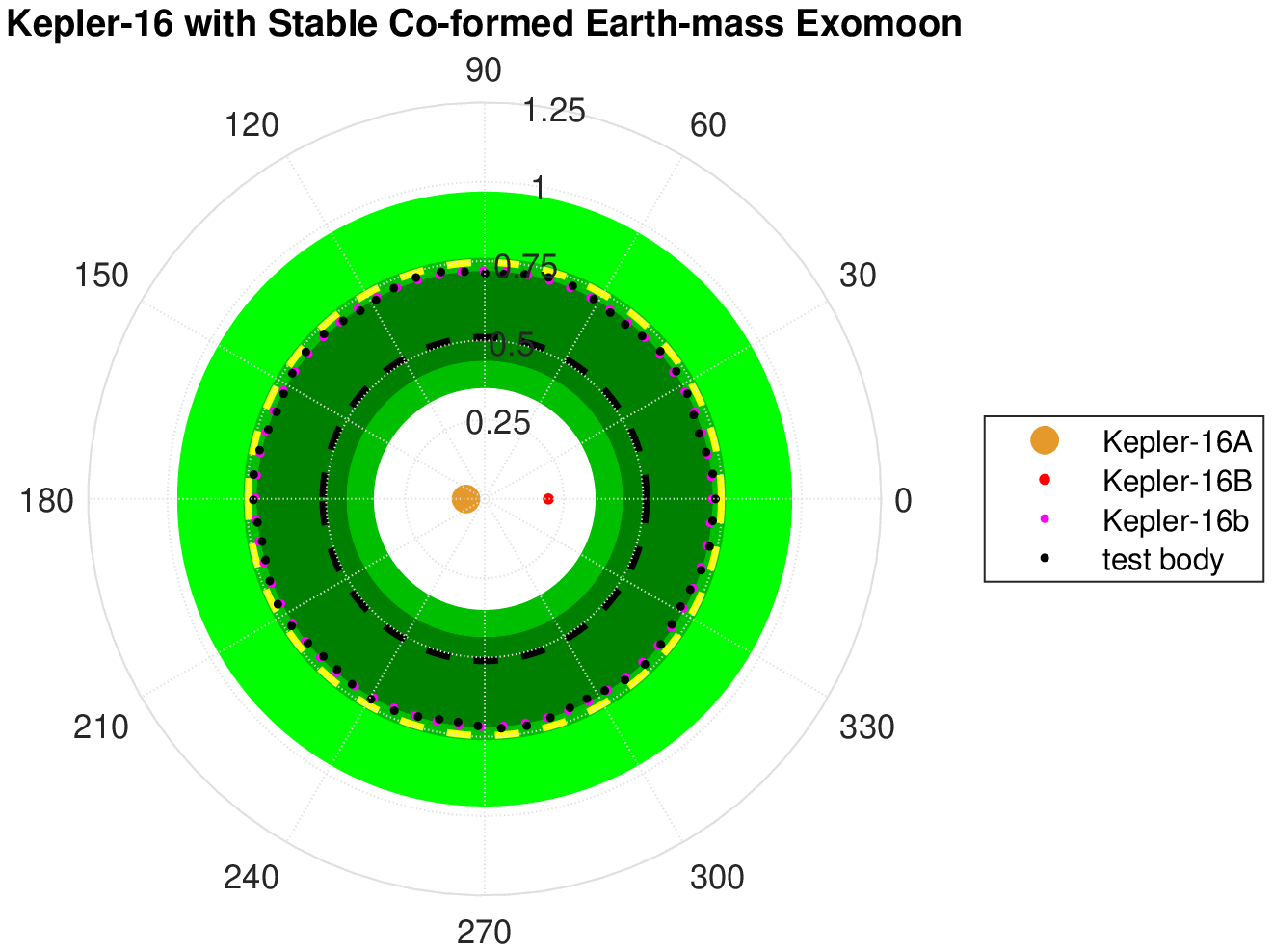,width=0.65\linewidth} \\
\noalign{\bigskip}
\noalign{\bigskip}
\epsfig{file=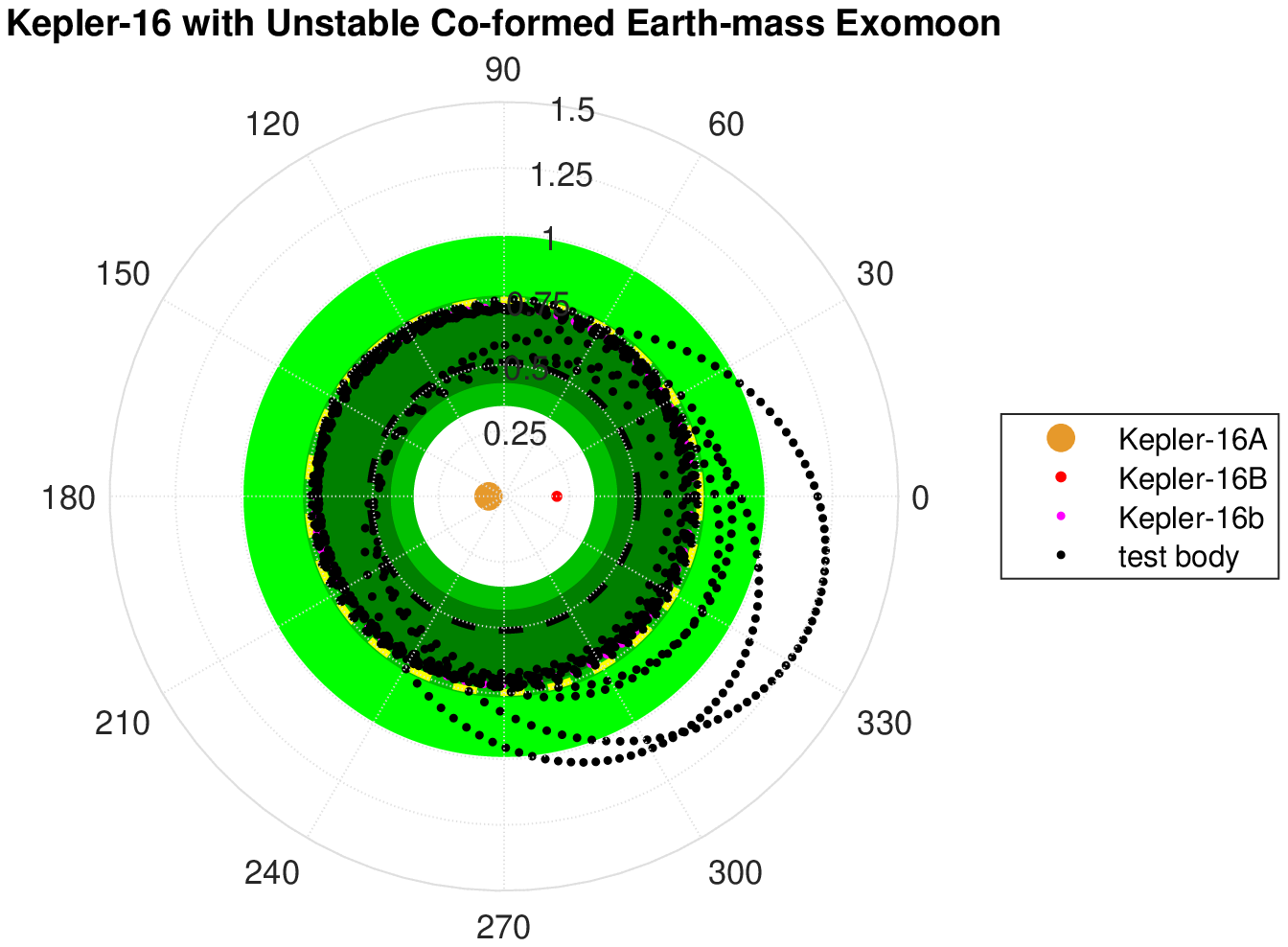,width=0.65\linewidth}
\end{tabular}
\caption{
Illustration of the previous results by QMC12 with updated HZ regions.
(a) Depiction of a stable S-type coformed Earth-mass exomoon
(black in QMC12); the primary and secondary stars (orange and red in QMC12,
respectively) and the Saturnian planet Kepler-16b are also given (magenta in
QMC12).  (b) Depiction of an unstable S-type coformed Earth-mass exomoon
(black in QMC12).  See Fig.~7 for information on the color coding of the HZs.
}
\end{figure*}

%+++++++++++++++++++++++++++++++++++++++++++++++++++++++++++++++++++++++

\clearpage

%%% *** Fig.9
%%%%%%%%%%%%%%%%%%%%%%%%%%%%%%%%%%%%%%%%%%%%%%%%%%%%%%%%%%%%%%%%%
\begin{figure*} 
\centering
\begin{tabular}{c}
\epsfig{file=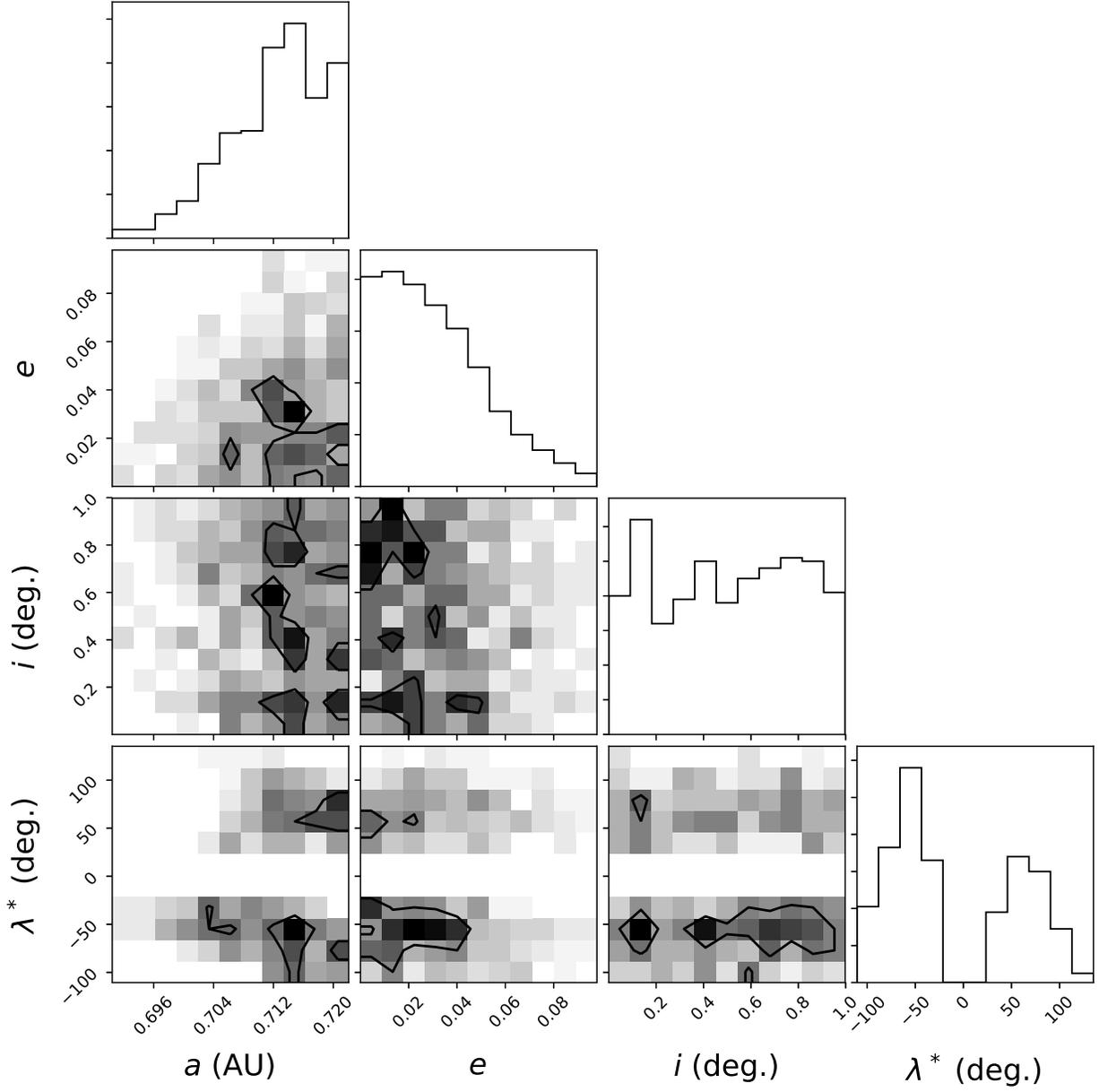,width=1.0\linewidth}
\end{tabular}
\caption{
Distributions of the initial semimajor axis $a$, eccentricity $e$,
inclination $i$, and relative mean longitude $\lambda^{\ast}$, given as
$\lambda^{\ast} = \lambda_{\oplus} - \lambda_{\rm 16B}$ that produces
a stable Earth-mass co-orbital planet in Kepler-16.  These initial conditions
are chosen relative to the center-of-mass of the stellar components. 
}
\end{figure*}

%+++++++++++++++++++++++++++++++++++++++++++++++++++++++++++++++++++++++

\clearpage

%%% *** Fig.10
%%%%%%%%%%%%%%%%%%%%%%%%%%%%%%%%%%%%%%%%%%%%%%%%%%%%%%%%%%%%%%%%%
\begin{figure*} 
\centering
\begin{tabular}{c}
\epsfig{file=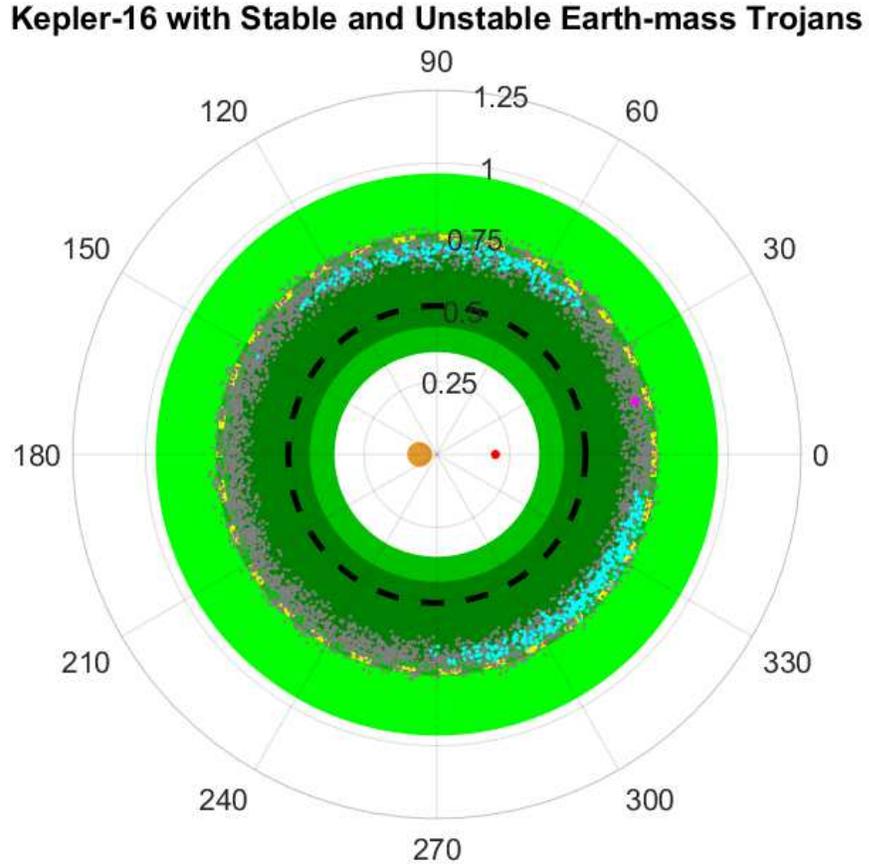,width=1.0\linewidth}
\end{tabular}
\caption{
Illustration of the starting locations of stable (cyan) and unstable (gray)
initial conditions out of 5,000 trials.  These simulations differ from
QMC12 as the relative phase between the binary and planetary orbit is now
taken into account, where the positive $x$-axis is taken to be the
line-of-sight.  See Fig.~7 for information on the color coding of the HZs.
}
\end{figure*}
%+++++++++++++++++++++++++++++++++++++++++++++++++++++++++++++++++++++++

%%% *** Fig.11
%%%%%%%%%%%%%%%%%%%%%%%%%%%%%%%%%%%%%%%%%%%%%%%%%%%%%%%%%%%%%%%%%
\begin{figure*} 
\centering
\begin{tabular}{c}
\epsfig{file=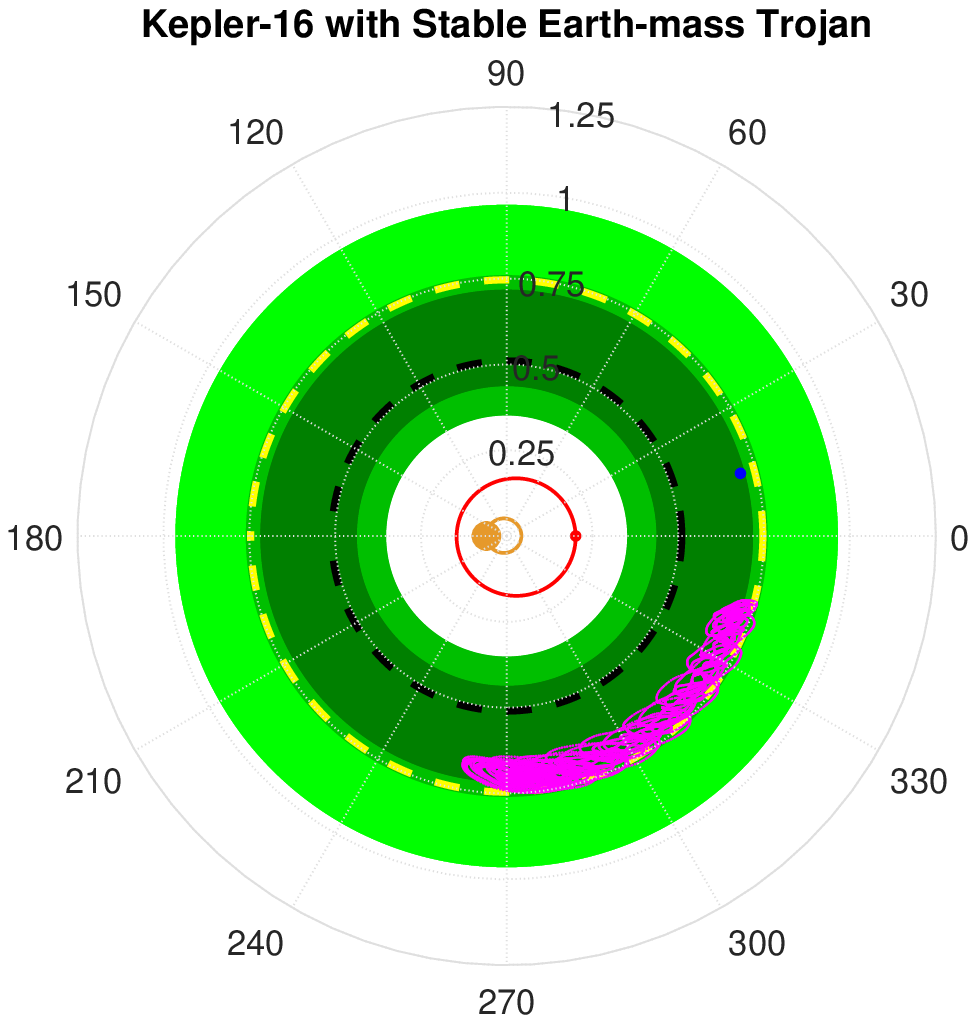,width=0.50\linewidth} \\
\noalign{\bigskip}
\noalign{\bigskip}
\epsfig{file=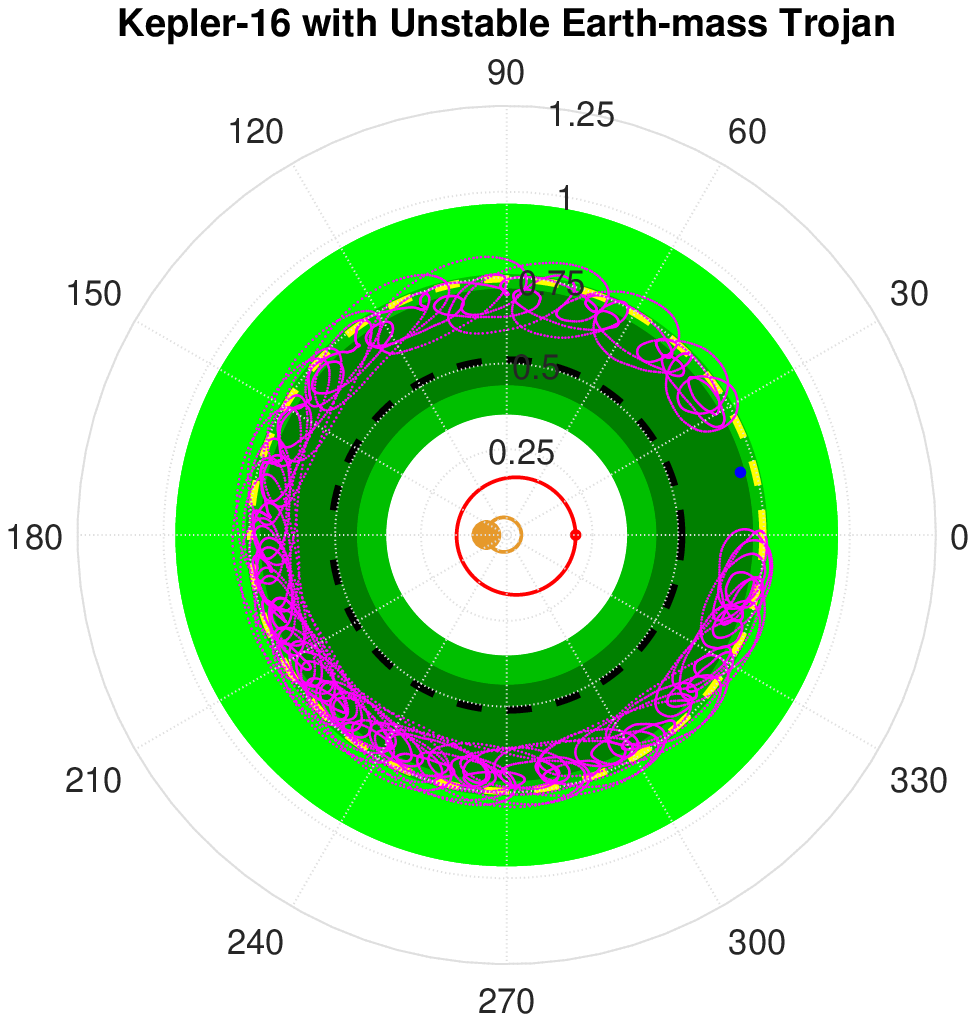,width=0.50\linewidth}
\end{tabular}
\caption{Examples of orbital evolution (magenta) of a stable (top) and
unstable (bottom) Earth-mass planet co-orbiting with Kepler-16b.
These orbits are shown in a rotated-reference frame depicting the
relative motions with Kepler-16b to illustrate Trojan (top) and
horseshoe (bottom) configurations.  See Fig.~7 for information on the
color coding of the HZs.
}
\end{figure*}

%+++++++++++++++++++++++++++++++++++++++++++++++++++++++++++++++++++++++

%%
%%  TABLES
%%

%+++++++++++++++++++++++++++++++++++++++++++++++++++++++++++++++++++++++

\clearpage

%%% *** Table 1
%%%%%%%%%%%%%%%%%%%%%%%%%%%%%%%%%%%%%%%%%%%%%%%%%%%%%%%%%%%%%%%%%%%%%%%%%
\begin{deluxetable}{lc}
\tablecaption{Stellar and Planetary Parameters of Kepler-16}
\tablewidth{0pt}
\tablehead{
Parameter   & Value$^a$
}
\startdata
Distance (pc)                           & $\sim$ 61                 \\       
$F_\mathrm{B} / F_\mathrm{A}$           & 0.01555 $\pm$ 0.0001      \\ 
$M_1$ $(M_\odot)$                       & 0.6897 $\pm$ 0.0035       \\  
$M_2$ $(M_\odot)$                       & 0.20255 $\pm$ 0.00066     \\   
$T_{\rm{eff},1}$ (K)                    & 4450 $\pm$ 150            \\
$T_{\rm{eff},2}$ (K)                    & 3308 $\pm$ 110            \\
$R_1$ $(R_\odot)$                       & 0.6489 $\pm$ 0.0013       \\
$R_2$ $(R_\odot)$                       & 0.22623 $\pm$ 0.00059     \\
$P_\mathrm{b}$ (d)                      & 41.079220 $\pm$ 0.000078  \\  
$a_\mathrm{b}$ (AU)                     & 0.22431 $\pm$ 0.00035     \\ 
$e_\mathrm{b}$                          & 0.15944 $\pm$ 0.00061     \\
$M_\mathrm{p}$ $(M_\mathrm{J})$         & 0.333 $\pm$ 0.016         \\
$a_\mathrm{p}$ (AU)                     & 0.7048 $\pm$ 0.0011       \\   
$e_\mathrm{p}$                          & 0.0069 $\pm$ 0.001        \\
\enddata
\tablecomments{
$^a$Data as provided by \cite{doy11} and reported by QMC12,
except for $T_{\rm{eff},2}$ and $R_2$, which have been determined
in this study.  All parameters have their usual meaning.
}
\label{table1}
\end{deluxetable}

%+++++++++++++++++++++++++++++++++++++++++++++++++++++++++++++++++++++++

\clearpage

%%% *** Table 2
%%%%%%%%%%%%%%%%%%%%%%%%%%%%%%%%%%%%%%%%%%%%%%%%%%%%%%%%%%%%%%%%%%%%%%%%%
\begin{deluxetable}{lc}
\tablecaption{Percentage Uncertainty of Kepler-16 Parameters}
\tablewidth{0pt}
\tablehead{
Parameter   & \% Uncertainty
}
\startdata
  $F_{B}/F_{A}$         & 0.64 \\
  $M_{1}$ ($M_{\odot}$) & 0.51 \\
  $M_{2}$ ($M_{\odot}$) & 0.33 \\
  $T_{\rm{eff},1}$ (K)  & 3.37 \\
  $T_{\rm{eff},2}$ (K)  & 3.33 \\
  $R_{1}$ ($R_{\odot}$) & 0.20 \\
  $R_{2}$ ($R_{\odot}$) & 0.26 \\
\enddata
\label{table2}
\end{deluxetable}

%+++++++++++++++++++++++++++++++++++++++++++++++++++++++++++++++++++++++

\clearpage

%%% *** Table 3
%%%%%%%%%%%%%%%%%%%%%%%%%%%%%%%%%%%%%%%%%%%%%%%%%%%%%%%%%%%%%%%%%%%%%%%%%
\begin{deluxetable}{lccc}
\tablecaption{Single Star Habitable Zone Limits}
\tablewidth{0pt}
\tablehead{
Habitable Zone Limit & Kas93/Und03  & Kop1314  & HZ Type
}
\startdata
Recent Venus              &  0.299  &  0.308  & RVEM (in)  \\
Runaway Greenhouse        &  0.334  &  0.390  & GHZ (in)   \\
Water Loss                &  0.376  &  0.402  & ...        \\
First CO$_2$ Condensation &  0.592  &  ...    & ...        \\
Maximum Greenhouse        &  0.708  &  0.723  & GHZ (out)  \\
Early Mars                &  0.746  &  0.766  & RVEM (out) \\
\enddata
\tablecomments{Kas93: \cite{kas93}, Und03: \cite{und03}, Kop1314:
\cite{kop13,kop14}
}
\label{table3}
\end{deluxetable}

%+++++++++++++++++++++++++++++++++++++++++++++++++++++++++++++++++++++++

\clearpage

%%% *** Table 4
%%%%%%%%%%%%%%%%%%%%%%%%%%%%%%%%%%%%%%%%%%%%%%%%%%%%%%%%%%%%%%%%%%%%%%%%%
\begin{deluxetable}{lcccc}
\tablecaption{Statistical Uncertainties}
\tablewidth{0pt}
\tablehead{
Habitable Zone Limit & \multicolumn{2}{c}{Kas93/Und03} & \multicolumn{2}{c}{Kop1314} \\
...                  & Min-Max & Statis & Min-Max & Statis
}
\startdata
Recent Venus              & 1.97 \% & 1.93 \% & 1.98 \% & 1.93 \% \\
Runaway Greenhouse        & 2.21 \% & 2.15 \% & 2.51 \% & 2.41 \% \\
Water Loss                & 2.51 \% & 2.44 \% & 2.59 \% & 2.51 \% \\
First CO$_2$ Condensation & 3.34 \% & 3.22 \% & ...     & ...     \\
Maximum Greenhouse        & 4.14 \% & 3.98 \% & 4.15 \% & 5.67 \% \\
Early Mars                & 4.40 \% & 4.26 \% & 4.39 \% & 5.97 \% \\
\enddata
\tablecomments{For references, see comments of Table~3.  Min-Max means that the
minimum/maximum values for the luminosities and effective temperatures are applied.
Statis means adequately applied statistical uncertainty propagation.
}
\label{table4}
\end{deluxetable}

%+++++++++++++++++++++++++++++++++++++++++++++++++++++++++++++++++++++++

\clearpage

%%% *** Table 5
%%%%%%%%%%%%%%%%%%%%%%%%%%%%%%%%%%%%%%%%%%%%%%%%%%%%%%%%%%%%%%%%%%%%%%%%%
\begin{deluxetable}{lccc}
\tablecaption{GHZ and RVEM RHZs of Binary System}
\tablewidth{0pt}
\tablehead{
Reference Distance      &   GHZ   & RVEM    & Relevance \\
...                     &   (AU)  & (AU)    & ...
}
\startdata
Inner RHZ Limit, innermost &  0.368  &  0.285  &  No   \\
Inner RHZ Limit, outermost &  0.444  &  0.361  &  Yes  \\
Outer RHZ Limit, innermost &  0.704  &  0.747  &  Yes  \\
Outer RHZ Limit, outermost &  0.783  &  0.827  &  No   \\
Orbital Stability Limit    &  0.510  &  0.510  &  Yes  \\
\enddata
\tablecomments{
The RHZ bounds have previously been referred to as RHLs \citep{cun14}.
Here the innermost and outermost points of these limits are reported,
which are of different relevance for setting the respective RHZ.
}
\label{table5}
\end{deluxetable}

%+++++++++++++++++++++++++++++++++++++++++++++++++++++++++++++++++++++++

\clearpage

%%% *** Table 6
%%%%%%%%%%%%%%%%%%%%%%%%%%%%%%%%%%%%%%%%%%%%%%%%%%%%%%%%%%%%%%%%%%%%%%%%%
\begin{deluxetable}{lc}
\tablecaption{EHZ of Kepler-16(AB)}
\tablewidth{0pt}
\tablehead{
$\epsilon$ &  EHZ   \\
...        &  (AU)
}
\startdata
2.0  &  0.765  \\
2.1  &  0.801  \\
2.2  &  0.837  \\
2.3  &  0.873  \\
2.4  &  0.910  \\
2.5  &  0.946  \\
2.6  &  0.982  \\
2.7  &  1.018  \\
2.8  &  1.055  \\
2.9  &  1.091  \\
3.0  &  1.127  \\
\enddata
%  \tablecomments{
%  }
\label{table6}
\end{deluxetable}

%+++++++++++++++++++++++++++++++++++++++++++++++++++++++++++++++++++++++

\clearpage

%%% *** Table 7
%%%%%%%%%%%%%%%%%%%%%%%%%%%%%%%%%%%%%%%%%%%%%%%%%%%%%%%%%%%%%%%%%%%%%%%%%
\begin{landscape}
\begin{deluxetable}{lcccccccc}
% \tabletypesize{\scriptsize}
\tablecaption{Comparison of Habitable Zone Limits}
\tablewidth{0pt}
\tablehead{
Type        &  \multicolumn{2}{c}{Single Star} & \multicolumn{6}{c}{Binary System Approach}          \\
\noalign{\smallskip}
\hline
\noalign{\smallskip}
 ... & GHZ  & RVEM & GHZ  & GHZ $(L^{-})$ & GHZ $(L^{+})$ & RVEM  & RVEM $(L^{-})$ & RVEM $(L^{+})$ \\
 ... & (AU) & (AU) & (AU) & (AU)          & (AU)          & (AU)  & (AU)           & (AU)
}
\startdata
RHZ$_{\rm in}$     & 0.390 & 0.308 & 0.444 & 0.419 & 0.470 & 0.361 & 0.341 & 0.381  \\
RHZ$_{\rm out}$    & 0.723 & 0.763 & 0.704 & 0.662 & 0.746 & 0.747 & 0.702 & 0.792  \\
$a_{\rm orb}$      & ...   & ...   & 0.510 & 0.510 & 0.510 & 0.510 & 0.510 & 0.510  \\
${\Delta}$HZ       & 0.333 & 0.455 & 0.194 & 0.152 & 0.236 & 0.237 & 0.192 & 0.282  \\
Type               & ...   & ...   &   PT  &   PT  &   PT  &   PT  &   PT  &   PT   \\
\enddata
\tablecomments{
$L^{+}$ and $L^{-}$ indicate $L \pm {\Delta}L$, respectively, with variations
in $T_{\rm{eff}}$ and $R$ simultaneously applied to both stellar components
(see Table~1).  ${\Delta}$HZ indicates the width of the HZ with consideration
of the orbital stability limit, if applicable.  PT conveys that the P-type HZ
is truncated due to the additional requirement of orbital stability.
}
\label{table7}
\end{deluxetable}
\end{landscape}

%+++++++++++++++++++++++++++++++++++++++++++++++++++++++++++++++++++++++

\clearpage

%%% *** Table 8
%%%%%%%%%%%%%%%%%%%%%%%%%%%%%%%%%%%%%%%%%%%%%%%%%%%%%%%%%%%%%%%%%%%%%%%%%
\begin{deluxetable}{llccccc}
\tablecaption{Initial Conditions for Exomoon Sample Cases}
\tablewidth{0pt}
\tablehead{
 Publication & Type        &  $a$ & $e$ & $i$        & $\omega$   & $M$    \\
             &             & (AU) &     & ($^\circ$) & ($^\circ$) & ($^\circ$)
}
\startdata
\multirow{6}{*}{\parbox{4cm}{QMC12}} & Kepler-16(AB) & 0.22431 & 0.15944 & 0 & 0 & 180\\
& Kepler-16(AB)b & 0.7048 & 0.0069 & 0 & 180 & 180\\
& Stable Retrograde & 0.619 & 0.13 & 180 & 180 & 180 \\
& Stable Trojan & 0.7048 & 0.0069 & 0 & 180 & 240 \\
& Stable Prograde & 0.715  & 0 & 0 & 180 & 180 \\
& Unstable Prograde & 0.721  & 0 & 0 & 180 & 180\\
\noalign{\smallskip}
\hline
\noalign{\smallskip}
\multirow{4}{*}{\parbox{4cm}{This Work}} & Kepler-16(AB) & 0.22431 & 0.15944 & 0 & 263.464 & $-171.114$\\
& Kepler-16(AB)b & 0.7048 & 0.0069 & 0.3079  & 318 & $-211.49$ \\
& Stable Trojan & 0.7096 & 0.0088  & 0.8175 & 37.499 & 35.272\\
& Unstable Trojan & 0.6902 & 0.0651 & 0.0795 & 124.235 & 154.849\\
\enddata
\tablecomments{
Initial conditions in terms of orbital elements for the binary (Kepler-16(AB)),
the Saturnian planet (Kepler-16b), and the possible Earth-mass exomoon.  These
orbital elements can be used to reproduce our new results (Fig. 11) and the
previous results of QMC12 (Figs. 7 and 8).
}
\label{table8}
\end{deluxetable}

% \clearpage

\end{document}